\documentclass[conference]{IEEEtran}
\IEEEoverridecommandlockouts
\usepackage{dblfloatfix}

\usepackage[numbers,sort&compress]{natbib}
\usepackage{amsmath,amssymb,amsfonts}
\usepackage{algorithm} 

\usepackage{algorithm}
\usepackage{algpseudocode}
\usepackage{algpseudocode} 
\usepackage{booktabs}
\usepackage{array}

\usepackage{pgfplots}
\pgfplotsset{compat=1.18}  
\usepackage{graphicx}
\usepackage{lipsum}
\usepackage{textcomp}
\usepackage{xcolor}
\usepackage{authblk}
\usepackage{pgf-pie}
\usepackage[utf8]{inputenc}
\usepackage{multirow}
\def\BibTeX{{\rm B\kern-.05em{\sc i\kern-.025em b}\kern-.08em
    T\kern-.1667em\lower.7ex\hbox{E}\kern-.125emX}}
\begin{document}




\title{Advanced Vulnerability Scanning for Open Source
Software: Detection and Mitigation of Log4j Vulnerabilities}

\author{Victor Wen and Zedong Peng}
\affil{University of Montana \\Missoula, MT,USA \\
victor.wen@umconnect.umt.edu, zedong.peng@mso.umt.edu
}

\maketitle
\begin{abstract}

Automated detection of software vulnerabilities remains a critical challenge in software security. Log4j is an industrial-grade Java logging framework listed as one of the top 100 critical open source projects. On Dec. 10, 2021 a severe vulnerability Log4Shell was disclosed before being fully patched with Log4j2 version 2.17.0 on Dec. 18, 2021. However, to this day about 4.1 million, or 33 percent of all Log4j downloads in the last 7 days contain vulnerable packages. Many Log4Shell scanners have since been created to detect if a user's installed Log4j version is vulnerable. Current detection tools primarily focus on identifying the version of Log4j installed, leading to numerous false positives, as they do not check if the software scanned is really vulnerable to malicious actors. This research aims to develop an advanced Log4j scanning tool that can evaluate the real-world exploitability of the software, thereby reducing false positives. Our approach first identifies vulnerabilities and then provides targeted recommendations for mitigating these detected vulnerabilities, along with instant feedback to users. By leveraging GitHub Actions, our tool offers automated and continuous scanning capabilities, ensuring timely identification of vulnerabilities as code changes occur. This integration into existing development workflows enables real-time monitoring and quicker responses to potential threats. We demonstrate the effectiveness of our approach by evaluating 28 open-source software projects across different releases, achieving an accuracy rate of 91.4\% from a sample of 140 scans. Our GitHub action implementation is available at the GitHub marketplace and can be accessed by anyone interested in improving their software security and for future studies. This tool provides a dependable way to detect and mitigate vulnerabilities in open-source projects.

\end{abstract}

\begin{IEEEkeywords}
Open Source Software, Vulnerability detection, Log4j, GitHub Action
\end{IEEEkeywords}

\section{Introduction}

Today, an overwhelming majority of codebases contain open source software (OSS). According to the Open Source Security and Risk Analysis (OSSRA) 2024 report, about 96\% of codebases across multiple industries use open source code, and among those an average of 77\% of the code in those repositories contain OSS \cite{Syno}. However, it is alarming that 84\% of these repositories that underwent risk assessment contain at least one open source vulnerability. Given the critical role of OSS, understanding its vulnerabilities becomes paramount. OSS is usually introduced into the code via direct download, or in many cases are installed as part of software ecosystems such as NPM, NuGet, and Maven. Although it is easy to update any outdated packages when any security patches are deployed, in many cases the organization may be unaware of any vulnerabilities in their software packages. Many organizations usually have software security governance rules in their pipeline, but those only get triggered when changes are made to specific repositories. Therefore vulnerabilities may not be picked up until months or even years later in the pipeline if not updates are made to the project. This issue may expand beyond the original repository as a project's dependencies may also contain vulnerabilities \cite{PKVIC}.

One method is to scan the codebase based on governance rules to detect any active Common Vulnerabilities and Exposures (CVEs) \cite{CVE}. These open source component governance rules can be set up and enabled for pipelines on Azure DevOPs, Maven Enforcer, or as an Github action. Component governance allows for the detection of any outdated or vulnerable packages which can be resolved by upgrading to a non-vulnerable version. However, sometimes developers may need to update other parts of their codebase to accommodate the upgrade as there may be some breaking changes if a major version bump is required. Refactoring the code could be time consuming as the developer would need to pinpoint the location of any breaking changes. OSS is often back-compatible but this can lead to other issues as well if an older class in the package containing vulnerabilities needs to be updated as well.

This brings us to the focus of this study—Log4j, an industrial-grade logging software developed by Apache. Log4j is particularly significant as it is used in over 8\% of the Maven ecosystem, which is the world's largest Java package repository. Critical open source projects like Log4j are considered ``critical" to the ecosystem based on several factors such as total downloads, contributors, and dependencies on the project. On Dec. 10, 2021, a severe vulnerability CVE-2021-44228, known as Log4shell, was disclosed to the public in which the JNDI lookup can be exploited to execute arbitrary code loaded from an Lightweight Directory Access Protocol (LDAP) server for Log4j2 \cite{apache}. This was fully patched by Apache on Dec. 17, 2021 but not before millions of devices were exploited. At the time the vulnerability was discovered in 2021, millions of devices were affected with over 35,000 Java packages impacted by Log4j vulnerabilities \cite{Goog}. 

While some codebases, such as Netty and MyBatis \cite{zhang2013research}, updated to Log4j2 version 2.17.0 on Dec. 19 and Dec. 18, 2021, respectively, many organizations struggled to apply the necessary patches for months. This was particularly true for those using Log4j V1, unsupported and deprecated since 2015 \cite{apache_v1}. Additionally, the average age of open source vulnerabilities in repositories is a concerning 2.8 years, with nearly half of these codebases inactive for the past two years\cite{Sona}. This indicates that both large and small organizations often lack the resources to address these vulnerabilities promptly, as they balance bug fixes with new development.

Moreover, the case of Mirth Connect by NextGen Healthcare highlights another aspect. As of the Log4Shell disclosure, Mirth Connect was using the unsupported Log4j version 1.2.16 \cite{mirth}, which had not received updates for over six years. Despite containing deprecated packages susceptible to various security risks, such as CVE-2021-4104 which allows data deserialization attacks, Mirth Connect was not vulnerable to CVE-2021-44228 due to the absence of the JndiLookup class in Log4j V1. Additionally, it was not exposed to CVE-2021-4104, as it did not use the JMSAppender class for logging \cite{mirth}. Mirth Connect only transitioned to the secure Log4j2 version 2.17.2 in Jul. 26, 2022.

This brings up a critical question: How many of these codebases marked with at least one OSS vulnerability are truly vulnerable? And how reliable are the CVE reports? For instance let's examine the braces package maintained by micromatch. Earlier this year, braces was exposed to CVE-2024-4068 (CVE score: 7.5), which fails to limit the number of characters the program can handle, leading to memory exhaustion. Despite the reported CVE, braces is not exploitable unless the user exploits it themselves locally as there is no way to submit any regular expressions via direct input \cite{micromatchCVE}. Therefore CVE-2024-4068 is not an actual security vulnerability and is at best just a performance boost after micromatch reviewed and tested the packages.

Our research aims to develop an advanced scanning tool that not only scans repositories via a GitHub action and local machines but also evaluates the real-world exploitability of vulnerabilities, such as Log4Shell, within those codebases. By leveraging GitHub actions, our tool provides automated and continuous scanning capabilities, ensuring that vulnerabilities are identified promptly as code changes are made, and offering mitigating solutions based on CVE reports. 

We show the effectiveness of our approach by evaluating it on 28 OSS projects publicly with publicaly available source code, achieving an accuracy rate of 91.4\%. The main contribution of our work is the development of an advanced scanning tool that not only detects vulnerable packages but also provides targeted recommendations for mitigating these detected vulnerabilities, providing new support for software security management. In what follows, we present background information in Section II. We then discuss the critical role of open-source software and the challenges in vulnerability detection in Section III. Section IV details our advanced scanning methodology, Section V describes the empirical evaluations, and finally, Section VI concludes the paper.

\section{Background}

In this sections, we will discuss the background of OSS and it's underlying security risks. Then we will examine the methodology and challenges to this project.

\subsection{Open Source Software}

 There are millions of open source projects, with a vast majority of these repositories hosted on sites such as Github. OSS is publicly developed and available for anyone to use it's source code and documentation under an open source license. There are many types of open source licenses such as MIT and Apache 2.0, they all have the same fundamental criteria. Projects utilizing an open source license must be publicly available for free, include the source code, as well as allow any modifications to the code to fit any purpose without discrimination \cite{OSI}. As such, the more popular OSS projects may have thousands of variations to fit an organization's use case. This flexibility can provide benefits such as cost reduction as well as security thanks to direct access to the source code along with increased code reusability \cite{PKVIC}. However, the widespread adoption and modification of OSS also present significant challenges \cite{stol2010challenges}. One major issue is the difficulty in maintaining security across multiple versions and variations of a project. With thousands of contributors and countless forks, vulnerabilities can be introduced and overlooked, leading to security risks that can propagate across different versions. Additionally, organizations may struggle to keep up with the latest security patches, especially when dependencies on multiple OSS projects are involved \cite{kula2018developers}. This issue is exacerbated by the fact that many organizations lack comprehensive tracking and management systems for OSS components, resulting in outdated and potentially vulnerable software remaining in use. These challenges underscore the importance of robust vulnerability detection and management tools to ensure the security and reliability of software that relies on open source components.

 \subsection{Security Vulnerabilities in Open Source Code}
The increased flexibility of OSS is a double edged sword since it is not only easier for 3rd parties to detect and report vulnerabilities, but also easier for bad actors to examine the source code to exploit them as well \cite{rajala2012strategic}. This is a concern since it is often faster to exploit publicly reported vulnerabilities than patching. Some OSS projects are not well maintained, leading to delays in deploying security patches to resolve security vulnerabilities. As open source projects are decentralized, it is up the the individual organizations to maintain the versioning within their codebase. However, many of these organizations have delays the patch from getting implemented due to limited developer resources. Developers not only have to patch software directly affected, but also project dependencies using vulnerable versions. 

The impact of projects with vulnerable dependencies cannot be understated as just one vulnerability can lead to large scale outages. On July 19th, a cybersecurity company called CrowdStrike released an update that resulted in a global IT outage and resulted in about 8.5 million windows devices facing the Blue Sceen of Death (BSOD) in what is the ``biggest IT outage in history" \cite{CrowdStrike}. Although this was not the result of an exploited vulnerability, this shows how potential impact security vulnerabilities in OSS can also cause a chain reaction if not patched and are exploited by bad actors.

  Let's examine the Log4Shell vulnerability for the Apache's Log4j2 package, CVE-2021-44228, which affected hundreds of millions of devices \cite{adam2022attack}. If we look at the official CVE report, we can see that it relates to the Log4Shell vulnerability in which attackers can use remote code execution (RCE) via Log4j2's JNDI feature \cite{CVE-2021-44228}. This gives hacker all the information they need to exploit software using vulnerable versions of Log4j as it specifically lists the affected feature and makes it easier for bad actors to focus attacks on the JDNI endpoint. This CVE is also an example of one of the challenges developers face when attempting to patch the vulnerability due to the popularity of the Log4j logging framework within Java based projects. Reusable libraries can have a security impact on any software that contains dependencies on the project \cite{PKVIC}: the Arduino IDE, Netty, MyBatis, and Elasticsearch are just a few examples of how other OSS projects dependent on Log4j were affected by CVE-2021-44228.
  
\subsection{Existing Vulnerability Detection Methods}
On the disclosure of the Log4Shell vulnerability, scanners were created to scan software for the packages affected. Some of the early versions created were very simple and only checked against the package versions such as log4j-checker \cite{log4jChecker}. More sophisticated scanners were later created to scan for Log4Shell in hosted applications such as log4j-scanner \cite{CISAgit} and these scanners mainly target HTTP-related ports by adding payloads to test for vulnerabilities 
\cite{hiesgen2022race}.

In addition to these basic scanners, more advanced detection methods have been developed. For example, Veracode’s platform provides static and dynamic application testing to identify vulnerabilities in both first-party code and open-source dependencies. Veracode also offers tools to monitor network traffic for suspicious Java class downloads, which can indicate exploitation attempts \cite{chen2020automated}. Furthermore, platforms like Datadog have integrated capabilities to detect Log4Shell exploit attempts by monitoring for suspicious Java class downloads and analyzing network traffic for unusual LDAP connections \cite{ZackChristophe}. While these scanners and tools are effective in identifying vulnerabilities, they also face challenges such as high false positive rates and the need for continuous updates to handle new variants of the exploit. Implementing a combination of these tools and maintaining an updated inventory of assets using Log4j can enhance the detection and mitigation of such vulnerabilities.

\subsection{Gaps in Current Detection Methods}

Current vulnerability scanners generally focus on identifying at-risk packages within a codebase by flagging outdated packages known to be vulnerable, such as those affected by Log4Shell. Alternatively, specialized tools like log4j-scanner aim to simulate payloads in the JNDI lookup string, flagging software as vulnerable if the simulated attack results in remote code execution \cite{CISAgit}.

However, these scanners have significant limitations. Firstly, they do not provide any insight into why specific codebases are exposed to vulnerabilities \cite{amankwah2017evaluation, chakraborty2021deep}. They primarily identify the presence of vulnerabilities but lack the capability to analyze and explain the underlying reasons for the exposure. This lack of detailed analysis makes it difficult for developers and security teams to understand the root causes of vulnerabilities and take appropriate remedial actions. Moreover, these tools often generate numerous false positives, flagging software as vulnerable based solely on the presence of specific package versions without considering the actual exploitability within the context of the software's environment.


\subsection{Github action}

GitHub Actions is a powerful CI/CD tool that facilitates the automation of building, testing, and deploying software in a continuous integration and continuous deployment pipeline \cite{kinsman2021software}. The steps for a GitHub action are typically configured in a YAML file located in the .github/workflows directory of a repository. These YAML files define the jobs to be executed when triggered by specific events, such as pushes, pull requests, or scheduled intervals. 

Previous studies have highlighted the impact and adoption of GitHub Actions in software development. For instance, Kinsman et al.\cite{kinsman2021software} examined over 3,000 repositories to investigate changes in various development activity indicators following the adoption of GitHub Actions. Valenzuela-Toledo and Bergel \cite{valenzuela2022evolution} conducted a study on the usage and maintenance practices of GitHub Actions workflows in popular GitHub repositories, identifying various types of workflow modifications. 
The primary advantage of using GitHub Actions for vulnerability scanning lies in its ability to provide immediate feedback to developers. As soon as a new code change is pushed to the repository, the GitHub action is triggered, running the configured security scans and reporting any detected issues. This enables developers to address vulnerabilities early in the development process, reducing the risk of security breaches in production environments. Similarly, we developed a customized GitHub action designed to automate the scanning of repositories for potential Log4j vulnerabilities. 


\section{Methodology}

\begin{figure*}
  \includegraphics[width=\textwidth]{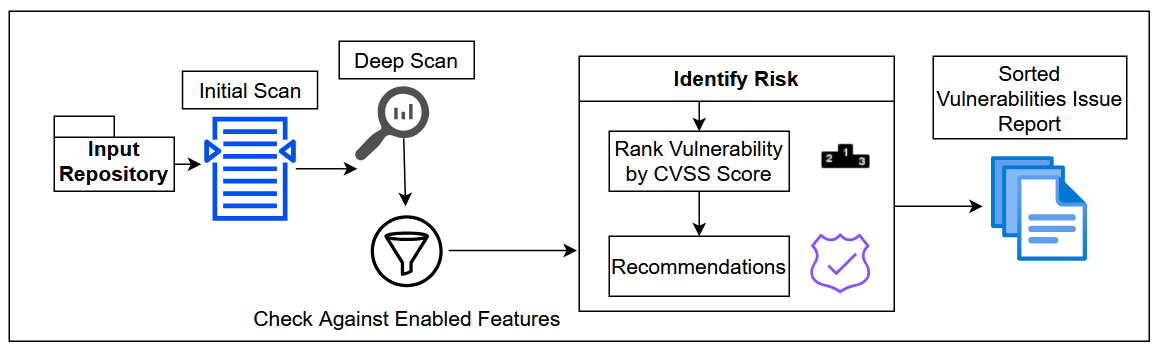}
  \caption{Scanner Flow}
  \label{fig:1}
\end{figure*}

In many cases, although the repository may contain vulnerable packages, they may not be exploitable. To fill this gap in scanner capability, we propose a scanner that analyzes the codebase itself. Given this goal, we examined 28 open-source repositories across hundreds of versions of projects, including Netty, Mirth Connect \cite{camacho2016automated}, MyBatis, and various Apache Foundation repositories with dependencies on Log4j. By using different releases of the same project, we can directly compare the vulnerable and patched versions of the repository against the results from our scanner. Figure~\ref{fig:1} provides an overview of our approach. There are three main steps in our methodology: initial scan, advanced deep scan, and risk identification through vulnerability ranking and validation.
\subsection{Collection of Open Source Software}
The first step in our methodology involves collecting a comprehensive dataset of open-source software repositories. This process includes:

\subsubsection{Selection Criteria}

\begin{itemize}
    \item \textbf{Dependency on Log4j}: We targeted repositories that have dependencies on Log4j to ensure relevant and meaningful data for our analysis.
    \item \textbf{Diverse Project Types}: A mix of projects from various domains and sizes, including popular projects like Netty, Mirth Connect, and MyBatis, as well as various Apache Foundation repositories.
    \item \textbf{Version History}: Repositories with extensive version histories were prioritized to allow for comparative analysis of different versions.
\end{itemize}

\subsubsection{Data Collection Process}

\begin{itemize}
    \item \textbf{GitHub API Usage}: We utilized GitHub's API to programmatically collect metadata and clone repositories that met our selection criteria.
    \item \textbf{Local Storage}: The collected repositories were cloned and stored locally to facilitate detailed analysis and repeated scanning.
\end{itemize}
\subsubsection{Initial Filtering}
\begin{itemize}
    \item \textbf{Initial Scan for Log4j Usage}: A preliminary scan was conducted to confirm the presence of Log4j dependencies in the collected repositories. This involved parsing project configuration files such as pom.xml for Maven projects to identify Log4j versions.
\end{itemize}

For each selected repository, a total of 5 releases were selected to be scanned for each project for a total of 140 total scans. Scanning the same number of releases for each repository preserves the consistency of the data collected as the result of each scan would hold the same weight for each repository. Another reason for only using 5 scans per project is due to the limitations due to the number of releases for Apache Ozone \cite{Ozone}. Apache Ozone only has 5 total releases in it's lifetime (1.0.0, 1.1.0, 1.2.1, 1.3.0, and 1.4.0) so limiting the total scans per project to 5 will ensure the weight of each scan is consistent for each repository.

\subsection{Initial Scan}



A preliminary scan was conducted to confirm the presence of Log4j dependencies in the collected repositories that are displayed on Algorithm 1. This process involved parsing project configuration files, such as pom.xml for Maven projects, to identify the versions of Log4j being used. The scan aimed to detect potentially vulnerable Log4j versions and prepare the repositories for a more in-depth analysis. The initial scan process can be summarized as follows:

\subsubsection{Parsing the Configuration File}
The \texttt{pom.xml} file is parsed to create a structured representation of the document.

\subsubsection{Extracting Dependencies}
The root element of the XML tree is examined to identify dependency elements.

\subsubsection{Identifying Log4j Versions}
For each dependency, the \texttt{artifactId} and version are extracted. If the \texttt{artifactId} contains ``Log4j," the version is checked against known vulnerable patterns:
\begin{itemize}
    \item Versions starting with ``1." are flagged as Log4j v1 vulnerabilities.
    \item Versions between 2.0 and 2.17.2 (excluding security patches 2.3.1 and 2.12.3) are flagged as Log4j v2 vulnerabilities.
\end{itemize}

\subsubsection{Handling Exceptions}
If parsing the \texttt{pom.xml} file fails or the \texttt{pom.xml} does not exist, the scanner will default to a deep scan of the entire repository.

This initial scan provides a baseline by identifying repositories that contain potentially vulnerable versions of Log4j. For instance, when scanning MyBatis version 3.5.16, the scan results in no vulnerabilities as it uses Log4j version 2.23.1. In contrast, MyBatis version 3.5.8 is flagged as vulnerable due to its use of a vulnerable Log4j version, triggering a deep scan of all files in the source folder. 

By conducting this initial scan, we filter out repositories without vulnerabilities, allowing us to focus our advanced analysis on those with potential security risks. This step ensures that our resources are used efficiently and that the deep scan process targets the most relevant codebases.

\begin{algorithm}
\caption{Check for Vulnerable Log4j Versions in pom.xml}
\begin{algorithmic}[1]
\Require file\_path: Path to the pom.xml file
\Ensure List of found vulnerabilities

\State \textbf{function} check\_pom\_file\_for\_vulnerable\_versions(file\_path)
    \State tree $\leftarrow$ Parse XML file from \textbf{file\_path}
    \State root $\leftarrow$ Get root element from \textbf{tree}
    \State found\_vulnerabilities $\leftarrow$ []

    \For{each dependency \textbf{in} root}
        \State artifactId, version $\leftarrow$ Extract from dependency
        \If{artifactId contains ``log4j"}
            \If{version starts with ``1."}
                \State Add Log4j v1 found to
                \textbf{found\_vulnerabilities}
            \ElsIf{version matches vulnerable pattern}
                \State Add Log4j v2 vulnerable to 
                \textbf{found\_vulnerabilities}
            \EndIf
        \EndIf
    \EndFor

    \State \Return found\_vulnerabilities
\State \textbf{end function}

\State \textbf{exception handling:} If parsing fails, print error and return []

\end{algorithmic}
\end{algorithm}

\begin{algorithm}
\caption{Deep Scan for Log4j Vulnerabilities}
\begin{algorithmic}[1]
\Require directory: Path to the directory to scan
\Ensure List of found vulnerabilities

\State \textbf{function} deep\_scan\_for\_log4j\_vulnerabilities(directory)
    \State found\_vulnerabilities $\leftarrow$ []
    
    \For{each file \textbf{in} directory and sub-directories}
        \If{file is not `pom.xml'}
            \State content $\leftarrow$ Read file content
            \State vulnerabilities $\leftarrow$ check\_log4j\_vulnerabilities(content)
            \State Append vulnerabilities to \textbf{found\_vulnerabilities}
        \EndIf
    \EndFor

    \If{\textbf{found\_vulnerabilities} is empty}
        \State \Return ``No vulnerabilities found."
    \Else
        \State \Return \textbf{found\_vulnerabilities}
    \EndIf
\State \textbf{end function}

\end{algorithmic}
\end{algorithm}
\subsection{Deep Scan}


If the initial scan detects vulnerable packages in the configuration file, the process progresses to a more thorough ``deep scan." The deep scan meticulously evaluates both the enabled and disabled features of Log4j within the software by searching for specific keywords and patterns throughout every file in the source code that is displayed on Algorithm 2. These patterns are defined based on the corresponding Common Vulnerabilities and Exposures (CVEs) and regular expressions that include critical information about specific Log4j features. If these features are enabled, they could potentially render the software exploitable by external threats. For this deep scan, we target specific patterns associated with known vulnerabilities. Specifically, all versions of Log4j v1 are flagged due to their inherent vulnerabilities. Additionally, all versions of Log4j v2 lower than version 2.17.2 are flagged, except for the security-patched versions 2.3.2 (for Java 6) and 2.12.4 (for Java 7). Beyond just identifying the version, the scanner is configured to check against a list of known vulnerable components within the Log4j library. This includes looking for specific classes and configurations that have been identified in CVEs as vectors for potential exploitation

\subsubsection{JNDI Lookup Class}
This pattern specifically looks for the org.apache.logging.log4j.core.lookup.JndiLookup class, which is associated with significant vulnerabilities like CVE-2021-44228 and CVE-2021-45046 \cite{apachelog4j}. CVE-2021-44228 allows attackers remotely execute code via LDAP servers \cite{CVE-2021-44228}, while CVE-2021-45046 enables malicious input data using a JNDI Lookup to perform remote code execution and Denial of Service attacks \cite{CVE-2021-45046}.
\subsubsection{SockerServer Class}
Looks for the presence of org.apache.log4j.net.SocketServer which is associated with CVE-2019-17571 \cite{apache_v1}. This vulnerability is affects Logj1 versions 1.2 up to 1.2.17, and the SockerServer class is vulnerable to deserialization of untrusted data and could be exploited via remotely excecuted code \cite{CVE-2019-17571}.
\subsubsection{SMTPAppender Class}
Identifies the use of the SMTPAppender class under org.apache.log4j.net.SMTPAppender linked to CVE-2020-9488 which was patched in Log4j 2.12.3 and 2.13.1. CVE-2020-9488 allows for SMTPS connections to be intercepted by man-in-the-middle attacks due to host mismatches with the Log4j2 SMTPAppender \cite{CVE-2020-9488}. This is caused by errors in the SslConfiguration and can result in the logs being leaked via the appender \cite{apachelog4j}.

\subsubsection{JMSAppender Class}
This pattern detects the use of the JMSAppender class org.apache.log4j.net.JMSAppender related to CVE-2021-4104. TopicBindingName and TopicConnectionFactoryBindingName configurations could be used by attackers to cause the JMSAppender to perform JDNI requests to perform remote code execution via manipulation of the LDAP \cite{CVE-2021-4104}. This vulnerability affects Logj1 and is related to the Log4Shell vulnerability.

\subsubsection{JMSSink Class}
Searches for the JMSSink Class org.apache.log4j.net.JMSSink connected to CVE-2022-23302 \cite{apache_v1}. This CVE is similar to CVE-2021-4104 in that it allows attackers to use JMSSink to perform JNDI requests that result in remote code execution due to deserialization of untrusted data. This exploit can only happen if the malicious actor has write access to configurations Log4j or references a LDAP service \cite{CVE-2022-23302}.

\subsubsection{JDBCAppender Class}
and lastly the JDBCAppender class which matches org.apache.log4j.jdbc.JDBCAppender associated with CVE-2022-23305. By default, the JDBCAppender accepts SQL queries as a configuration and almost always uses the \%m message converter \cite{apache_v1}. As such, this leaves the JDBCAppender class open to SQL injection attacks which allow attackers to execute unintended SQL scripts \cite{CVE-2022-23305}

By analyzing the interaction between active configurations and known vulnerabilities, the deep scan can accurately assess the actual risk posed by the software. Following this thorough analysis, the software is then flagged as either ``not vulnerable'' or ``vulnerable''. By employing this deep scan methodology, we can accurately identify potential vulnerabilities that extend beyond simple version checks. This thorough evaluation is crucial for understanding the actual risk posed by these vulnerabilities and sets the stage for the subsequent steps of our methodology: CVE ranking, recommendation, and manual validation. These next steps involve quantifying the severity of identified vulnerabilities and ensuring their accuracy, thereby enabling developers to prioritize and address the most critical security issues effectively.

\subsection{CVE Ranking}
After the deep scan identifies potential vulnerabilities, each one is ranked according to its severity using the Base Score from the Common Vulnerability Scoring System (CVSS) \cite{4042667}. The CVSS Base Score ranges from 0 to 10, with 10 representing the most severe vulnerabilities. This score provides a standardized assessment of a vulnerability's inherent qualities, taking into account factors such as exploitability and impact. Exploitability considers elements like the attack vector, complexity, and required privileges, while impact assesses the potential effects on the system's confidentiality, integrity, and availability if the vulnerability is exploited. 
The CVSS Base Score is calculated as follows:

\begin{equation}
\text{Base Score} = \lceil \text{Impact} + \text{Exploitability} \rceil
\end{equation}

where:

\begin{itemize}
  \item \textbf{Impact:} Represents the overall effect of a successful exploit on the target system.
  \item \textbf{Exploitability:} Reflects how easy it is for an attacker to exploit the vulnerability.
\end{itemize}

By using the CVSS Base Score, we provide an objective measure of each vulnerability's criticality, enabling developers to prioritize their remediation efforts effectively. This ensures that attention is focused on the vulnerabilities that pose the greatest risk to the system's security, facilitating a more efficient and targeted approach to vulnerability management.

\subsection{Recommended Actions and Report Generation}

In addition to flagging vulnerable features and marking each with it's related CVE number and CVSS Base Score, the report will also provide the user with useful information regarding potential steps to take to mitigate the vulnerability including the minimum Log4j version needed to patch along with steps that can be taken to resolve the security risks by updating configurations or disabling the offending features.

Depending on the CVEs detected, the scanner will provide the following recommended actions:
        \subsubsection{``CVE-2021-44228:"}
        Upgrade to Log4j 2.17.1 or later. If upgrading is not possible, mitigate by setting the system property log4j2.formatMsgNoLookups \cite{log4jConfig} to true or removing the JndiLookup class from the classpath.
        \subsubsection{``CVE-2021-45046"}: Upgrade to Log4j 2.17.0 \cite{apache} or later. This vulnerability is an extension of CVE-2021-44228 and requires an update to mitigate the risks effectively \cite{CVE-2021-45046}. This vulnerability could also be mitigated by removing the JndiLookup class.
        \subsubsection{``CVE-2021-45105"}: Upgrade to Log4j 2.17.0 or later. This vulnerability is related to uncontrolled recursion from self-referential lookups. Updating the configurations to either remove references to Context Look up or replacing it with Thread Context Map patterns will address this vulnerability \cite{apache}.
        \subsubsection{``CVE-2021-44832"} Upgrade to Log4j 2.17.1 or later. This vulnerability affects Log4j 2.0-beta9 to 2.17.0 and involves a remote code execution vulnerability due to improper configuration \cite{CVE-2021-44228}. JDBCAppender could be set to only accept JNDI data sources from java protocols as well if upgrade is not currently possible \cite{apache}.
        \subsubsection{``CVE-2019-17571"} Upgrade to Log4j 2.8.2 or later. This vulnerability affects Log4j 1.x and allows deserialization of untrusted data, leading to remote code execution \cite{CVE-2019-17571}. Can mitigate by deleting SocketServer.class from the jar.
        \subsubsection{``CVE-2020-9488"} Upgrade to Log4j 2.13.2 or later. This vulnerability involves a misconfiguration that could lead to a denial of service or remote code execution \cite{CVE-2020-9488}. This host mismatch vulnerability can be mitigated by setting mail.smtp.ssl.checkserveridentity to true to enable hostname validation \cite{apache}.
        \subsubsection{``CVE-2021-4104"} This affects Log4j 1.x versions. Since Log4j 1.x is no longer supported, the best course of action is to upgrade to Log4j 2.x. If upgrading is not possible, ensure that the JMSAppender is not configured in your logging configuration files \cite{apache_v1}.
        \subsubsection{``CVE-2022-23302"} Upgrade to Log4j 2.17.1 or later. This vulnerability allows a remote attacker to execute code via a crafted input and can be mitigated by disabling or removing JMSSink \cite{CVE-2022-23302} from the configurations.
        \subsubsection{``CVE-2022-23305"} Upgrade to Log4j 2.17.1 or later. This vulnerability allows for a denial of service (DoS) through uncontrolled recursion and should be mitigated by updating or removing the JDBCAppender from the configuration.
        \subsubsection{``CVE-2022-23307} Upgrade to Log4j 2.17.1 or later. This critical vulnerability allows for remote code execution and should be addressed immediately. Can mitigate by removing dependencies to the Apache Chainsaw \cite{CVE-2022-23307} project from the configuration.
        \subsubsection{``Potential misconfiguration"} Remove any unneeded appenders and ensure all configurations are up to date. 

        Developers can prioritize immediately mitigating the found vulnerabilities by disabling affected features or removing vulnerable classes and appenders from the Log4j configuration file. This would ensure projects are unable to be exploited by the detected CVEs until official security patches are releases, and will be explored in more detail in the Evaluation section of this paper. The scanner will also generate a report of all the vulnerable enabled features that can be exploited and the user can then take steps to either update the packages to a non-vulnerable version, or disable certain features by eliminating feature bloat to remove any unused or at risk code.
       
\subsection{Github Action for Scanner}
The next step was to create a publicly available Github action to easily scan any Github repository. We created a Github action called Log4jDeepScanAction which could be easily integrated into the CICD pipeline. This will checkout the repository under ``\$GITHUB\_WORKSPACE" using the existing checkout action \cite{checkout} which allows our workflow access and run against the repository. This enables developers to check against the scanner in real time as changes are made since the output from the scanner will show if the codebase is still vulnerable after patching. Our GitHub action implementation is available at GitHub marketplace (https://github.com/marketplace/actions/log4j-vulnerability-scanner).

\subsection{Validation and Challenges}
For this project, we collected and downloaded the source code of the selected repositories and release versions locally to scan against the script. This is because OSS is not exclusively stored on Github and thus it is not always possible to utilize the Github action except for the latest release when available. The results for each are then recorded and compared against official release notes, issues, and documentation. For example, we installed and scanned MyBatis v3.5.16 (the latest version), v3.5.8 and v3.5.9. Based on the release notes the scanner should find v3.5.9 and v3.5.16 to be non-vulnerable while flagging MyBatis v3.5.8 \cite{myBatis}. This is because we can see that the official issues list and release note state that MyBatis was upgraded to Log4j 2.17.0 with v3.5.9. However, the scanner reports that MyBatis v3.5.9 is vulnerable to CVE-2021-45105 which can cause Denial of Service \cite{CVE-2021-45105} but based on the release notes, this version should no longer be vulnerable as MyBatis could be used without Log4j and the pom.xml configurations for Log4j are optional. Scanning against different release versions allow for the comparison of the effectiveness of the scanner to official release notes of patches to Log4j vulnerabilities. 

During the ranking process, we prioritize using the Base Score from the CVSS to rank vulnerabilities in open-source projects. While the CVSS framework also includes Temporal and Environmental Scores, obtaining accurate assessments for these metrics presents significant challenges in the context of open-source software \cite{4042667,gifford2009temporal,walkowski2021vulnerability,mell2007complete}.

The Temporal Score is intended to reflect the current state of a vulnerability, taking into account factors such as exploit availability, remediation level, and report confidence \cite{mell2007complete}. However, the dynamic nature of open-source projects makes it difficult to maintain accurate assessments of these factors. Open-source projects often undergo rapid changes, with frequent updates and patches, making it challenging to assess the remediation level consistently. 
Similarly, assessing the Environmental Score accurately is problematic due to the diverse deployment environments of open-source software \cite{mell2007complete}. These projects can be deployed across numerous configurations and security measures, making it challenging to generalize an environmental assessment that applies to all users. 
These contextual details are often unknown or unavailable when analyzing open-source projects, making the Environmental Score difficult to ascertain.

\section{Evaluation}
In this section, we examine the effectiveness of our scanner in detecting vulnerable Log4j packages and features. In addition, we will also analyze the reports generated to validate the accuracy of the scanner results against official release notes. In our evaluation of the advanced vulnerability scanning tool, we focused on analyzing the effectiveness and accuracy of the scanner against various open source repositories. These repositories were chosen based on their relevance and history with the Log4j vulnerability, ensuring a comprehensive assessment across different versions and configurations.

\begin{table*}
\centering
\caption{List of repositories with their respective information}
\begin{tabular}{|c|c|c|c|c|c|}
\hline
\textbf{Repository name} & \textbf{Ref.} & \textbf{Author} & \textbf{Release} & \textbf{Last Updated} & \textbf{Vulnerable Versions} \\ \hline
Apache Spark & \cite{spark} & Apache & May 26, 2014 & Apr 18, 2024 & Apache Spark: 1.0.0 - 3.2.0 \\ \hline
Arduino IDE & \cite{arduino} & Arduino & Nov. 30, 2011 & Feb. 20, 2024 & Arduino IDE: before 1.8.18 \\ \hline
Apache Hive & \cite{hive} & Apache & Jan 11, 2013 & May 20, 2024 & Apache Hive: 0.6.0 - 3.1.2 \\ \hline
Apache Wicket & \cite{wicket} & Apache & Jun, 2005 & Jun 17, 2024 & Apache Wicket 8.13.0 and lower
\\ \hline
MyBatis-3 & \cite{myBatis} & MyBatis & May 19, 2010 & Apr 4, 2024 & MyBatis-3:5.8 and lower \\ \hline
Netty & \cite{netty} & Netty & 2004 & Jul 19, 2024 & 4.1.72 and lower \\ \hline
Mirth Connect & \cite{mirth} & NextGen HealthCare & Jul 18, 2006 & Jul 27, 2024 & 4.0.0 and lower \\ \hline
Elasticsearch & \cite{elastic} & Elastic & Feb 10, 2010 & Mar 11, 2024 & lower than 7.16.1 / 6.8.21 \\ \hline
Log4j2 & \cite{apachelog4j} & Apache & Jan 8, 2013 & Mar 10, 2024 & 2.0-beta7 through 2.17.0 excluding 2.3.2 and 2.12.4 \\ \hline
phoss-smp & \cite{phoss} & phax & Aug. 7, 2016 & May 24, 2024 & phoss SMP 5.5.0 and lower \\ \hline
Apache Pulsar & \cite{pulsar} & Apache & Aug. 31, 2016 & Jun. 5, 2024 & Apache Pulsar: 2.8.2 and lower \\ \hline
Apache Tapestry & \cite{tapestry} & Apache & Apr. 2002 & Apr. 16, 2024 & Apache Tapestry: 5.0 - 5.7.3 \\ \hline
Apache Nifi & \cite{nifi} & Apache & Jul. 26, 2015 & Jul. 1, 2024 & Apache Nifi: 0.1 - 1.15.0 \\ \hline
Apache Traffic Control & \cite{traffic} & Apache & Jan. 22, 2018 & Apr. 3, 2024 & Apache Traffic Control: 1.1.2 - 6.0.1 \\ \hline
Apache SkyWalking & \cite{skywalking} & Apache & Dec. 30, 2015 & May 29, 2024 & SkyWalking: 3.0.3 - 8.9.0 \\ \hline
Apache OFBiz-Framework & \cite{OFBiz} & Apache & Apr. 2010 & May 2024 & OFBiz: 4.0 - 18.12.02 \\ \hline
Apache JMeter & \cite{jmeter} & Apache & Dec. 15, 1998 & Jan. 9, 2024 & Apache JMeter: 2.0 - 5.4 \\ \hline
Apache Jena & \cite{jena} & Apache & Aug. 28, 2000 & Jul. 12, 2024 & Jena: 2.7.1 - 4.3.0 \\ \hline
Apache Geode & \cite{geode} & Apache & Oct. 15, 2016 & May, 2024 & Apache Geode: 1.12.0 - 1.14.0 \\ \hline
Apache Fortress & \cite{fortress} & Apache & Apr. 15, 2015 & Sep. 6, 2023 & Apache Fortress: 2.0.6 \\ \hline
Apache Druid & \cite{druid} & Apache & Jun. 18, 2014 & Jun. 16, 2024 & Apache Druid: before 0.22.1 \\ \hline
Apache Calcite Avatica & \cite{calcite} & Apache & Nov. 6, 2013 & May 6, 2024 & Apache Calcite Avatica: 1.10.0 - 1.19.0 \\ \hline
Apache Archiva & \cite{archiva} & Apache & Nov. 2005 & Mar. 20, 2023 (Retired) & Apache Archiva: 1.0 - 2.2.5 \\ \hline
Apache Tika & \cite{tika} & Apache & Mar. 27, 2007 & Jul. 15, 2024 & Apache Tika: 1.0 - 2.2.0 \\ \hline
Apache Solr & \cite{solr} & Apache & Dec. 22, 2006 & May 29, 2024 & Apache Solr: 7.4.0 - 8.11.0 \\ \hline
Apache Flink & \cite{flink} & Apache & Aug. 26, 2014 & Mar. 18, 2024 & Apache Flink: 1.10.0 - 1.14.0 \\ \hline
Apache EventMesh & \cite{eventmesh} & Apache & Aug. 20, 2020 & Dec. 19, 2023 & Apache EventMesh: 1.0.0 - 1.2.0 \\ \hline
Apache Ozone & \cite{Ozone} & Apache & Sep. 2, 2020 & Jan. 19, 2024 & Apache Ozone: 1.0.0 - 1.2.0 \\ \hline 
\end{tabular}
\label{table:repos}
\end{table*}


\subsection{Input repositories}
For the evaluation of our advanced vulnerability scanning tool, we selected a diverse set of open-source repositories known for their relevance and history with Log4j vulnerabilities. This comprehensive assessment includes various projects from the Apache Foundation and other significant open-source contributors. All our experimental materials are publicly available at https://doi.org/10.5281/zenodo.13188600.

Table~\ref{table:repos} summarizes 28 repositories analyzed, providing key details such as repository names, authors, release dates, last updated dates, and specific versions identified as vulnerable to Log4j. These repositories were chosen based on several key factors. First, each repository has a documented history of vulnerabilities associated with Log4j, providing a rich dataset for testing the effectiveness of our scanner. Second, the repositories include both medium and large-sized projects, ensuring a comprehensive assessment of the scanner's effectiveness across different software complexities. A significant portion of the repositories, 21 out of 28, are classified as large projects \cite{winters2020software}, each exceeding 100,000 lines of code (LOC). These large projects, such as Apache Spark and Elasticsearch, represent substantial and complex systems commonly used in enterprise environments. 7 out of 28 repositories are classified as medium-sized projects, with LOC ranging from 20,000 to 100,000 \cite{winters2020software}. Examples of these medium projects include MyBatis-3 and Log4j2, which are widely used frameworks that offer a balance between complexity and manageability.

The selected repositories span a wide range of release dates and last updated dates, indicating both the longevity and current activity of these projects. For instance, Apache Spark \cite{spark}, with its initial release in 2014 and last update in 2024, represents a well-maintained project with ongoing updates. Similarly, the inclusion of older projects like Apache JMeter \cite{jmeter}, first released in 1998, highlights the long-term relevance of Log4j vulnerabilities across various software generations.

By comparing the scanner's detection capabilities with documented vulnerabilities and official release notes, we aim to validate its accuracy and reliability.

\subsection{CVE Rank and Analysis}
Table \ref{tab:cve_scores} details the ranking of various CVEs associated with Log4j and other related software, ordered by their CVSS scores. CVE-2021-44228 and CVE-2022-23307 \cite{CVE-2021-44228} \cite{CVE-2022-23307}, both scoring 10.0, and are identified as the most severe vulnerabilities, emphasizing their critical impact on system security. CVE-2022-23307 is related to a deserialization issue related to CVE-2020-9493, in which the Apache Chainsaw project has Java deserialization that could lead to could lead to malicious code execution \cite{CVE-2020-9493}. This affects all Apache Chainsaw versions under 2.1.0, and since Log4j1 contains a dependency on Apache Chainsaw versions under 2.0.0, it is also affected. Other notable CVEs, such as CVE-2021-45046 and CVE-2022-23302, also receive high scores above 9.0, underscoring the significant risks they pose. 

\begin{figure*}[!t]
\centering
\includegraphics[width=\textwidth]{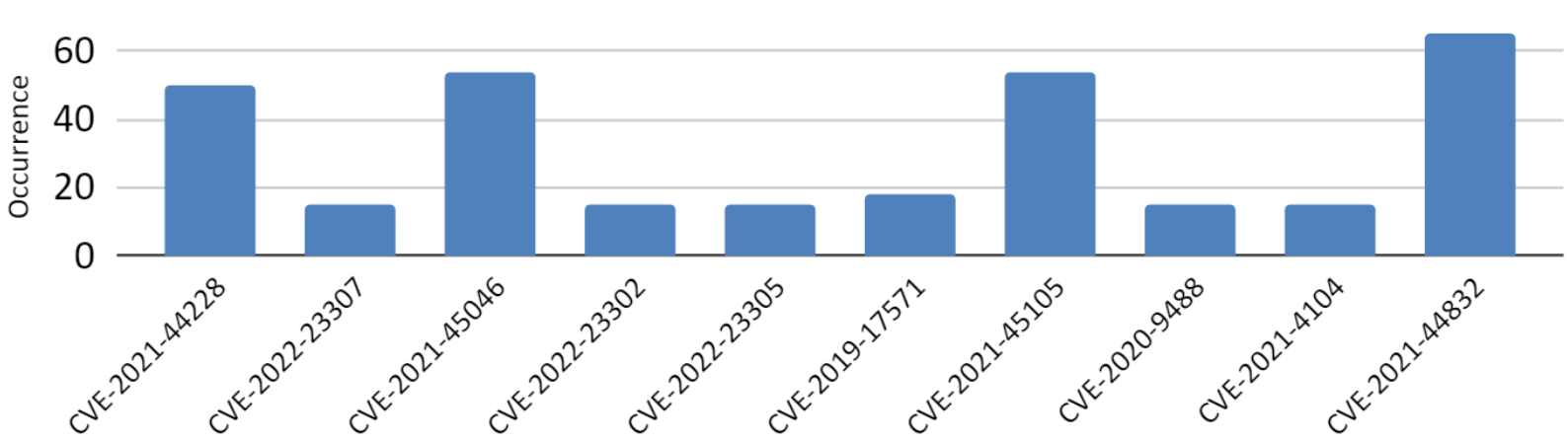} 
\caption{CVEs Occurrence}
\label{fig:CVEs Occurrence}
\end{figure*}

Our scanning of 140 repositories is displayed in Figure \ref{fig:CVEs Occurrence}, and shared more detailed experimental materials at https://doi.org/10.5281/zenodo.13188600. we found significant occurrences of several critical vulnerabilities. Notably, CVE-2021-44228 and CVE-2021-45046, both highly critical vulnerabilities with CVSS scores of 10.0 and 9.0 respectively, were found in 50 and 54 repositories, underlining their widespread exploitation risk. Conversely, CVE-2021-44832, despite a lower CVSS score, appears in 65 repositories, indicating a potentially underrecognized threat. The lower occurrence of CVE-2022-23307, found in only 15 repositories, might suggest less exploitation in the wild, or possibly more effective mitigation strategies already in place within these environments.

\begin{table}[ht]
    \centering
    \caption{CVE Scores Sorted by Score}
    \begin{tabular}{>{\raggedright\arraybackslash}p{5cm}r}
        \toprule
        \textbf{CVE Identifier} & \textbf{Score} \\
        \midrule
        CVE-2021-44228 & 10.0 \\
        CVE-2022-23307 & 10.0 \\
        CVE-2021-45046 & 9.0 \\
        CVE-2022-23302 & 9.0 \\
        CVE-2022-23305 & 9.1 \\
        CVE-2019-17571 & 9.8 \\
        CVE-2021-45105 & 7.5 \\
        CVE-2020-9488 & 7.5 \\
        CVE-2021-4104 & 7.5 \\
        CVE-2021-44832 & 6.6 \\
        Potential misconfiguration & 5.0 \\
        \bottomrule
        \\
    \end{tabular}
    \label{tab:cve_scores}
\end{table}

\subsection{Results and Recommendations}

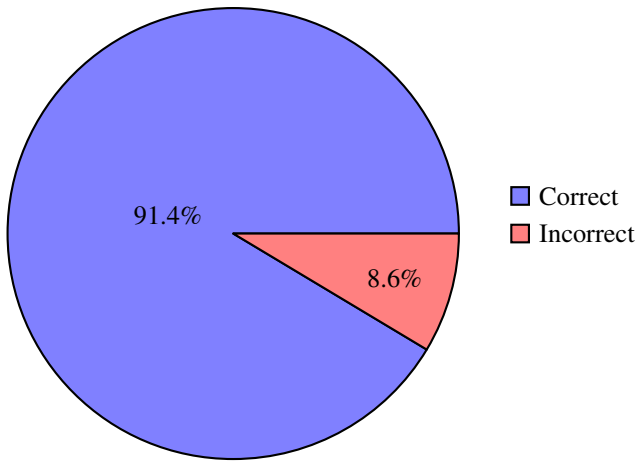
\begin{figure}[ht]
    \centering
    \begin{tikzpicture}
        \pie[
            color={blue!50, red!50},
            text=legend,
            radius=3
        ]{
            91.4/Correct, 8.6/Incorrect
        }
    \end{tikzpicture}
    \caption{Distribution of Correct and Incorrect Scans}
    \label{fig:scans_pie}
\end{figure}


The analysis of the patch release dates for the Log4Shell vulnerability (CVE-2021-44228) \cite{CVE-2021-44228} shows significant variation in response times among different open-source projects. Most projects addressed the vulnerability within a week, with an average patch time of approximately 5.83 days, excluding Apache Spark, Apache Traffic Control, Apache Wicket, and Mirth Connect, as those projects have dependencies on Log4j 1.x instead of Log4j 2.x. The evaluation of the Log4Shell patch timelines reveals key insights into the effectiveness of vulnerability management across different projects. This rapid response aligns with the methodological emphasis on the need for swift action in mitigating security risks associated with widely used libraries like Log4j. Projects such as Elasticsearch and Apache SkyWalking demonstrated exemplary readiness by releasing patches within a day or two of the disclosure, underscoring the importance of having established security protocols and efficient deployment processes in place.

This evaluation also highlights disparities in response times, with some projects like Arduino IDE and Apache Tapestry taking 12 and 11 days respectively to implement the patch for Log4Shell. This suggests potential challenges such as resource constraints, the need for extensive testing, or project-specific complexities that can delay patch releases. One reason for this delay is partly because the initial Log4Shell patch Log4j 2.15.0 was found to be insufficient in non-default configurations, so the JNDI class was disabled by default in 2.16.0 to fix this issue \cite{apachelog4j}. Moreover, the coordinated efforts observed within the Apache Software Foundation, where multiple projects released patches in quick succession, reflect the benefits of strong internal communication networks and shared resources, which were discussed as critical components in the proactive management of OSS security risks. 

During the evaluation, the scanner was applied to 5 releases of each repository, particularly focusing on versions known to have vulnerabilities associated with Log4j as well as the latest releases which are known to be non-vulnerable. The results were then cross-referenced with official release note documentation and CVE databases to validate the scanner's findings. 

The accuracy of our approach is displayed in Figure \ref{fig:scans_pie}. Out of a total of 140 scanned repositories, a total of 128 reported accurate results and correctly flagged the repository as either ``not vulnerable" or ``vulnerable", while 12 of the scans resulted in the scanner incorrectly flagging the repository as ``vulnerable", ``not vulnerable", or flagged incorrect CVEs linked to found vulnerabilities. The report generated by the scan is validated against the following: the known release version in which Log4j vulnerabilities were patched, Github issues list, and release notes. The report generated provides the developer with the related CVE, the affected package versions found, as well as the file path where the package reference is detected. 

\begin{figure*}

  \includegraphics[width=\textwidth]{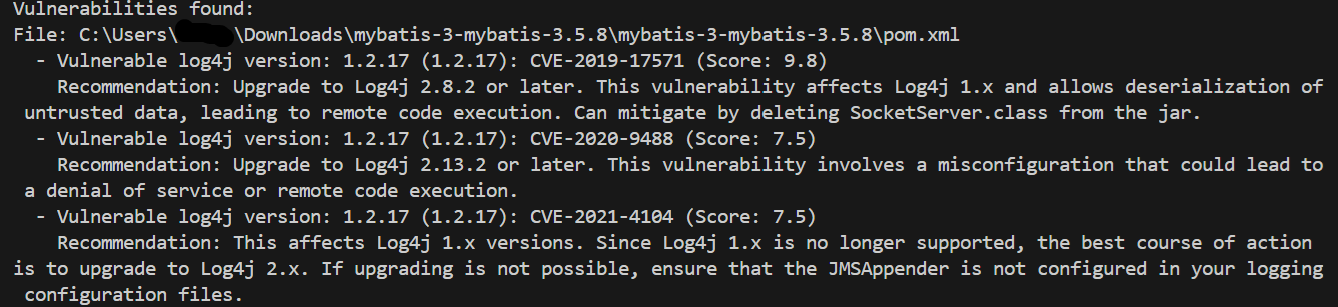}
  \caption{Recommended Security Mitigations for MyBatis Vulnerabilities}
  \label{fig:Mybatis}
\end{figure*}

Finally, the scanner also provides recommendations based on the detected vulnerable Log4j versions as seen in our methodology. For instance, as illustrated in Figure \ref{fig:Mybatis}, the scanner offers specific recommendations after scanning MyBatis. The developer can then take action to mitigate any security risks by disabling or removing affected classes as a temporary measure until the official security patches are released and prevents any malicious actors from exploiting the vulnerability. This means repositories can be secured quickly by disabling features and developers using the affected OSS do not need to wait for the official patch. If this scanner had been available on the day the CVE was published, more codebases could have been secured quickly by updating to configurations to mitigate the exploits. Such actions include setting log4j2.formatMsgNoLookups to true, effectively disabling the point of attack, and removing affected classes such as the JNDILookup, 

After the initial response to Log4Shell, the speed of releases upgrading Log4j versions slowed after CVE-2021-44228 and CVE-2021-45046 were patched since the remaining CVEs related to Log4j 2.x were lower in severity. Majority of the projects analyzed in this project did not upgrade to Log4j 2.17.1 or 2.17.2 (the current minimum non-vulnerable versions \cite{apachelog4j}) until after the second quarter of 2022.

\subsection{Scanner Erroneous Results}
Of the 12 scan reports were found to contain inaccurate information after user validation, none of them were false-negatives, and a total of 7 resulted in false-positive results and 5 others flagged the repository for incorrect CVEs. Out of the 7 false positive results, all 7 contained residue Log4j 1.x dependencies still within the source code. These included scans for Mirth Connect, Apache Spark, and MyBatis. The scan for MyBatis 3.5.9 was due to the scanner being unable to determine that Log4j was now optional after that release, and is no longer the default logger for MyBatis. Furthermore, Log4j 1.x was also fully deprecated in 3.5.9, and subsequent scans on 3.5.10 and higher resulted in no Log4j vulnerabilities found.

The other 5 erroneous reports were due to incorrect labeling of CVEs. One such case is Apache Pulsar's release 2.9.1 and 2.9.2 which were upgraded to Log4j 2.16.0 and 2.17.1 respectively. However, although these releases had patched their source code to resolve the Log4Shell vulnerability, they both had dependency on Netty 4.1.72Final which was still using Log4j 2.15.0 \cite{netty} \cite{pulsar}. This is an example of an inaccurate scan, as one of the drawbacks to this scanner is it is currently only able to scan against files and folders within the source but is unable to check against other external dependencies directly. We detected this vulnerability during the validation stage, as Apache Pulsar's release notes mention upgrades to Netty 4.1.72Final for 2.9.1 and 2.9.2, but we had confirmed that 4.1.72Final release of Netty was vulnerable on prior scans and analysis of Netty's releases.

\section{Threats to Validity}
A potential threat to construct validity in our study arises from the methodology used to determine the presence of vulnerabilities. The scanner's recommendations are based on detected vulnerable Log4j versions, relying on configurations that might not cover all possible real-world scenarios. While our tool aims to provide accurate results by cross-referencing CVEs with detected vulnerabilities, there may be discrepancies between the scanner's output and actual system configurations in diverse environments. This could impact the recall and precision of our findings, as the recommendations are contingent on specific configurations that might not be universally applicable.

The internal validity of our evaluation is supported by the rigorous manual assessment conducted by two researchers, which involved 30 hours of detailed analysis. This manual evaluation ensured that the scanner's detection capabilities were accurately assessed against known vulnerabilities and official release notes. However, there is a threat that subjective judgment during manual evaluation might influence the outcomes, although steps were taken to minimize bias by having multiple researchers independently verify the findings. 

A threat to external validity is the generalizability of our results beyond the 28 open-source repositories included in our study. While these repositories represent a range of sizes and complexities, they may not fully capture the diversity of all open-source projects potentially affected by Log4j vulnerabilities. 

\section{Future Works and Conclusion}
In this paper, we presented a GitHub Action-based scanner designed to detect and mitigate Log4j vulnerabilities, providing actionable insights and reducing false positives. Our evaluation across 140 scans across 28 open-source repositories demonstrated an accuracy of 91.4\% in identifying critical CVEs such as CVE-2021-44228 and CVE-2021-45105, along with timely recommendations for mitigation. Our automated approach is available in the GitHub Marketplace \cite{Log4jDeepScanAction}. 

Our work can be extended towards several avenues. Expanding the scanner’s applicability to other programming languages and platforms will enhance its utility across a broader range of software projects. Natural Language Processing (NLP) can be used to analyze unstructured data from commit messages, issue reports, and security bulletins to automatically identify potential risks and prioritize them based on severity.

\bibliographystyle{ieeetr}
\bibliography{bibi}
\end{document}